\documentclass[twocolumn,showpacs,preprintnumbers]{revtex4}
\usepackage{graphicx}% Include figure files
\usepackage{dcolumn}% Align table columns on decimal point
\usepackage{bm}% bold math

\begin{document}

%\preprint{APS/123-QED}
\preprint{}

\title{X-ray views of neutron star low-mass X-ray binaries}

\author{Sudip Bhattacharyya}
 \affiliation{Department of Astronomy and Astrophysics, Tata Institute of Fundamental Research, Mumbai 400005, India}%Lines break automatically or can be forced with \\
 \email{sudip@tifr.res.in}
%\affiliation{%

%\affiliation{


\begin{abstract}
A neutron star low-mass X-ray binary is a binary stellar system with a
neutron star and a low-mass companion star rotating around each other. In this
system the neutron star accretes mass from the companion, and as this
matter falls into the deep potential well of the 
neutron star, the gravitational potential energy is released primarily
in the X-ray wavelengths.
Such a source was first discovered in X-rays
in 1962, and this discovery formally gave birth to the ``X-ray astronomy".
In the subsequent decades, our knowledge of these sources has increased enormously by the
observations with several X-ray space missions. Here we give a brief overview
of our current understanding of the X-ray observational aspects of these
systems. 
\end{abstract}

\maketitle

\section{Introduction}

Cosmic objects emit in diverse electromagnetic 
wavelengths from radio to $\gamma$-rays. Some
of them (e.g., stars) are luminous primarily in a narrow wavelength range,
while others (e.g., active galactic nuclei) emit in a broad energy spectrum.
It is, therefore, essential to observe a celestial source in various
wavelengths in order to fully understand its nature. Up to the middle of the
twentieth century, visible light was the primary window for observation.
But in the last few decades, wavelength-based branches of astronomy have 
not only been developed, but also got matured. Moreover, many 
new types of sources have been 
discovered by the observations in wavelengths other than visible lights.
One such discovery was made in X-rays using rocket-borne instruments in 1962
\citep{Giacconietal1962}. These instruments detected X-rays from sources
outside the solar system for the first time, and specifically discovered a new type
of source Scorpius X-1. This discovery brought a new era in astronomy,
and formally gave birth to a new branch called ``X-ray astronomy".
The new source Scorpius X-1 was the first observed X-ray binary, and more 
specifically it was the first detected low-mass X-ray binary system 
containing a neutron star and a low-mass stellar companion.

Unlike the optical (i.e., visible light) and radio detection instruments, the X-ray
instruments must be sent above the atmosphere for observations. This is
because X-rays from space cannot reach the surface of the Earth. In 1960's,
balloons and sub-orbital rockets were extensively used to lift
the X-ray instruments. However, as these carriers
are short-lived, satellites were thought to be ideal for 
X-ray observations. The first astronomy satellite {\it Uhuru} was launched
by NASA in 1970. It identified over 300 discrete X-ray sources.
NASA's {\it Einstein} satellite (launched in 1978) was the first fully imaging X-ray 
telescope put into space. Its few arcsecond angular resolution increased
the sensitivity enormously, and made the imaging of extended objects, 
and the detection of faint sources possible. 
The {\it ROSAT} satellite launched in 1990 by the Germany/US/UK collaboration
did the first X-ray all-sky survey using a highly sensitive imaging telescope,
and detected more than 150000 objects.
The scientific outcome of these key X-ray space missions, as well as the 
other past X-ray satellites, have revolutionized our knowledge of the X-ray sky.

The current primary X-ray space missions are (1) NASA's {\it Rossi X-ray Timing Explorer}
({\it RXTE}); (2) European Space Agency's {\it XMM-Newton}; (3) NASA's {\it Chandra}; and
(4) Japan's {\it Suzaku}. They primarily operate in the $\approx 1-10$ keV range
(i.e., soft X-rays), and contain a variety of instruments.
The primary instrument of {\it RXTE} is a set of five
proportional counters (called ``Proportional Counter Array" or PCA) with large
collecting area and very good time resolution ($\approx 1$ microsecond). Therefore,
PCA is an ideal instrument to study the fast timing phenomena, as well as the continuum
energy spectrum. However, it is not an imaging instrument, and due to poor energy
resolution, PCA is not very suitable to study spectral lines. This instrument 
operates in $2-60$ KeV band, although the effective collecting area becomes small
above $\approx 20$ keV.
Another very useful instrument of {\it RXTE} is ``All-Sky Monitor" or ASM. It consists
of three relatively low-sensitivity proportional counters with large field of view,
and hence monitors the entire sky to detect transient X-ray sources. {\it XMM-Newton}
contains X-ray telescopes, charge-coupled devices and high resolution 
``Reflection Grating Spectrometers (RGS)". Therefore, this satellite,
that operates in $\approx 0.2-12$ keV range, is a good
imager, and is ideal for spectral analysis.
The {\it Chandra} observatory contains an X-ray telescope, charge-coupled devices,
high resolution camera and transmission gratings. Its angular resolution ($\sim$ 
subarcsecond) is the best among all the past and current X-ray instruments,
which makes it a unique imager. {\it Chandra} is also ideal to detect and 
study narrow spectral lines, and it operates in $\approx 0.1-10$ keV.
Finally, the {\it Suzaku} satellite also has X-ray telescopes and charge-coupled devices.
Since it contains an additional ``Hard X-ray Detector" (HXD), {\it Suzaku} can study the broadband 
energy spectrum.

All these X-ray space missions, especially the current ones, have gathered
a wealth of information about the low-mass X-ray binary systems, and hence our understanding of 
these sources has grown enormously. In this review,
we will give a brief overview of the X-ray observational aspects of neutron star 
low-mass X-ray binaries.
We note that the reference list will not be exhaustive in this short review.

\section{Neutron Stars}
%Give details and extreme physics of neutron stars. Describe various neutron
%star systems.

%No figure.

Neutron stars are extremely compact objects with a mass of about one
solar mass ($M_{\odot}$), but a radius of $\sim 10$ km. The core density of 
neutron stars are $5-10$ times the nuclear density, and their magnetic
field can be anywhere between $10^7-10^{15}$ Gauss. They can spin with 
a wide range of frequency values: very slow to several hundred Hertz.
Study of these stars provides a unique opportunity to probe
some extreme aspects of the universe, that cannot be achieved by other 
means.

Although the neutron stars were discovered in 1967, their plausible
existence was predicted in 1930's. Around this time,
Subrahmanyan Chandrasekhar calculated the upper limit 
of the mass of a collapsed star (white dwarf) that is supported
by the electron degeneracy pressure against the gravity. This limit
is the celebrated ``Chandrasekhar mass limit", with a value of 
about $1.4 M_{\odot}$. If the mass of the collapsed star is
greater than this limit, then the star is expected to collapse more.
But if the mass is less than a certain value,
the neutron degeneracy pressure will eventually balance the gravity to form a
stable star, known as neutron star \citep{BaadeZwicky1934, OppenheimerVolkoff1939}. 
This remained as a theoretical 
calculation almost for three decades, after which Hewish et al. \citep{Hewishetal1968} 
reported the discovery of a 1.377 sec period radio pulsar at
81.5 MHz, and Gold \citep{Gold1968, Gold1969} showed that the pulsars are 
spinning neutron stars with large surface magnetic fields.

The density of a neutron star increases by $\approx 14$ orders of magnitude
from the surface to the centre. As a result, the constituents and their 
properties change dramatically with the radial distance (see, for example, 
Shapiro \& Teukolsky \citep{ShapiroTeukolsky1983}). 
The dense core of a neutron star may contain pion condensates, deconfined
quarks, or other exotic (and possibly unknown) forms of matter, which makes 
the study of this part of the star extremely interesting. Moreover
according to a model, some of the neutron stars are actually ``strange stars"
made entirely of u, d and s quarks \citep{Bagchietal2006}.
Unfortunately, in spite of many theoretical studies, the properties of matter in 
the core of a neutron star is not yet known, because such a dense degenerate matter 
cannot be created in the laboratory.
Perhaps the only way to understand the nature of this high density matter
is to constrain the theoretically proposed equation of state models of 
the stellar core. This can be done by measuring three independent structural
parameters (e.g., mass, radius, spin frequency) of the same neutron star,
because the stable stellar structure gives a single mass vs. radius curve
for a known spin frequency and a given equation of state model \citep{LattimerPrakash2007}. 
Measurement of the stellar parameters is, therefore,
a primary goal of the neutron star astronomy. Such a measurement,
especially of the radius, is an extremely formidable task. This can be
appreciated from the fact that we need to measure the $\sim 10$ km radius
of a neutron star with $< 5$\% error, while the typical distance of the 
star is $\sim 10^{17}$ km.

Neutron stars can emit in various wavelengths depending on their properties and
the systems they are in. They are observed as radio, X-ray and 
$\gamma$-ray pulsars (that may create pulsar wind nebulae), millisecond
pulsars in binary systems, magnetars, and so on, and can be found in 
supernova remnants, X-ray binaries, etc. In this review we will concentrate
on neutron stars in X-ray binaries.

\section{Low-mass X-ray Binary}
%Evolution, other description, etc.

%Figure Roche lobe: got from Dipankar.
%Figure LMXB got from Dany Page.

\begin{figure}[btp]
\includegraphics[width=3.00in]{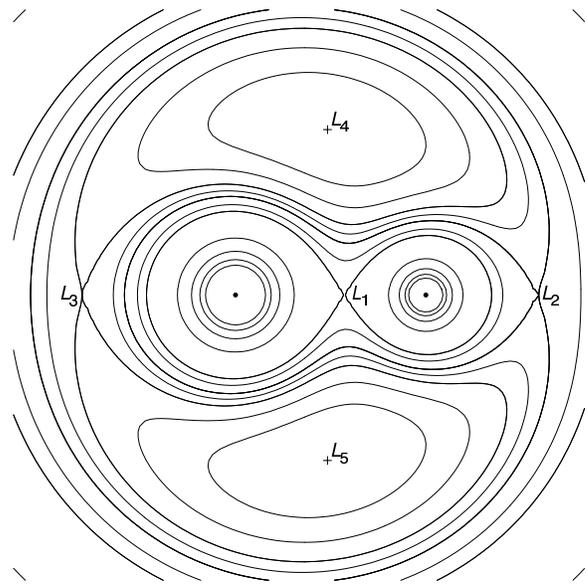}
\caption{Sections in the orbital plane of the equipotential surfaces of a 
binary system with the mass ratio 2:1. The two gravitating bodies are rotating
around each other, and the above mentioned potential includes the effects of 
both gravitational force and centrifugal force. The two parts of the figure-of-eight
equipotential (containing the $L_1$ point) are called Roche lobes, and the inner Lagrange 
point $L_1$, having a saddle-like potential, allows the matter to pass from one lobe
to the other (figure courtesy: Dipankar Bhattacharya).
}
\end{figure}

\begin{figure}
\includegraphics[width=3.50in]{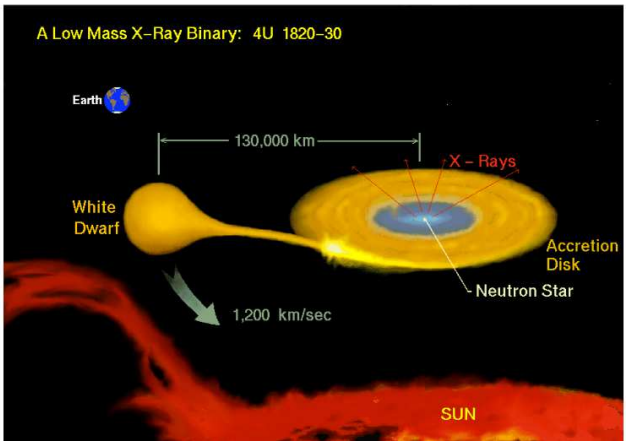}
\caption{Artist's impression of a neutron star LMXB 4U 1820-30 \citep{Liuetal2007}.
This is a particularly small LMXB with the orbital period of 11 minutes
(Sun and Earth are shown for an easy comparison of size).
The companion is naturally a white dwarf star, 
because only such a star can have the size similar to that
of companion's Roche lobe in these small systems (figure courtesy: 
Dany P. Page).
}
\end{figure}

A low-mass X-ray binary (LMXB) is an old binary stellar system (typical age 
$\sim 10^9$ years) with a low-mass companion star
($\leq 1M_{\odot}$; main sequence star, evolved star or white dwarf) and
a primary compact star (neutron star or black hole) rotating around each other. In such
a system, the companion star fills its Roche lobe, i.e., the critical equipotential
surface (see Fig. 1), and matter from it flows towards the compact star. 
Because of the initial angular momentum, this matter cannot move radially 
towards the primary star, and follows slow spiral paths that form an 
accretion disk. In this review, we consider the systems with neutron stars as
the primary stars. In such neutron star LMXBs, the accreted matter 
eventually hits the stellar surface, and generates electromagnetic radiation.
The accretion disk also emits such radiation by viscous dissipation, and
both these emissions are powered by the gravitational potential energy
release. The inner part of the accretion disk, as well as the neutron star
surface, emit X-rays, as their typical temperature is $\sim 10^7$ K.
Fig. 2 gives an artist's impression of such a system.
In the next section, we will give the basic theory of a simple geometrically
thin accretion disk in Newtonian gravity. Note that a more realistic
approach may include complicated fluid dyanamical processes
(e.g., Mukhopadhyay 2006 \citep{Mukhopadhyay2006}; Mukhopadhyay et 
al. \citep{Mukhopadhyayetal2005}) and general relativistic 
effects (e.g., Mukhopadhyay \& Misra 2003 \citep{MukhopadhyayMisra2003}). 
Moreover, the magnetosphere of the 
neutron star can affect the accretion process depending on the accretion 
rate and the magnetic field strength. Ghosh et al. \citep{Ghoshetal1977}  
and Ghosh \& Lamb \citep{GhoshLamb1978}
provided important theoretical understanding of this process.
Due to accretion induced angular momentum transfer, the neutron stars
in LMXBs are expected to be spun up. This is consistent with the 
observed stellar spin frequencies (typically $\sim 300-600$ Hz).
The magnetic fields of these neutron stars are relatively low 
($\approx 10^7-10^9$ Gauss) compared to the expected fields of these stars
when they are born, and several groups have worked on this problem of
stellar magnetic field evolution \citep{RayKembhavi1985, Srinivasanetal1990,
JahanMiriBhattacharya1995, Konaretal1995, ChoudhuriKonar2002, 
KonarChoudhuri2004}. Note that the low magnetic
field allows the accretion disk to approach very close to the neutron star.
This, in combination with the fact that X-rays can be detected both from the
star and the inner part of the disk for many LMXBs, provides an excellent opportunity
to study the extreme physics of and around neutron stars. However,
the modeling of any observation of the neutron star vicinity should
consider the special and general relativistic effects including light
bending (e.g., Datta \& Kapoor \citep{DattaKapoor1985}).

LMXBs are observed mostly in the disk, bulge and globular clusters of our galaxy.
Some of them have also been discovered from the nearby galaxies,
and the locations much above our galactic disk plane. According to the 
catalogue of Liu et al. \citep{Liuetal2007}, 187 LMXBs (containing black hole or 
neutron star) were known from our galaxy, 
``Large Magellanic Cloud" and ``Small Magellanic Cloud" up to that time.
LMXBs emit primarily in X-ray wavelengths, and the optical to X-ray
luminosity ratio is normally less than 0.1 for them.
The central X-ray emission from these sources cannot be spatially resolved 
using the current X-ray imaging instruments, because the distances 
of even our galactic LMXBs are very large (typically $\sim 10^{17}$ km).
Therefore, in order to understand the neutron star LMXBs, we have to 
rely on their spectral and timing properties. In this review, we will
describe some of these properties, as well as the information they provide.

\subsection{Theory of Accretion Disk}

In a neutron star LMXB, the companion mass is transferred to the neutron star
typically via a geometrically thin accretion disk. Therefore,
in this section we will briefly discuss the basic theory of such a disk.
Here we note that the standard model of steady thin accretion disks in
Newtonian gravity was worked out by Shakura \& Sunyaev \citep{ShakuraSunyaev1973}. 

In a thin accretion disk using cylindrical coordinates ($r, \theta, z$), we 
expect that the $\theta$ component $v_{\theta}$ should be the main component
of velocity, and the $z$ component $v_z = 0$. The accreted matter should 
move around the neutron star with a nearly Keplerian speed (i.e., the angular
speed $\Omega \approx \sqrt{GM/r^3}$; $M =$ mass of the neutron star), and a 
small radial inflow (speed $= v_r << v_{\theta}$) caused by the effect of viscosity
should exist. We consider $\partial/\partial\theta = 0$ because of the 
$\theta$-symmetry of the system.
Then the equation of continuity is
\begin{eqnarray}
{\partial\Sigma\over{\partial t}} + 
{1\over{r}} {\partial\over{\partial r}}(r\Sigma v_r) = 0,
\end{eqnarray}
where $\Sigma$ is the surface density of the accretion disk.
Using this and the $\theta$-component of the Navier-Stokes equation, we get
\begin{eqnarray}
{\partial\over{\partial t}}(\Sigma r^2 \Omega) + {1\over{r}} {\partial\over{\partial r}}
(\Sigma r^3 \Omega v_r) = \Lambda,
\end{eqnarray}
where $\Omega = v_{\theta}/r$ and $\Lambda$ is the term involving viscosity.
We note that $2\pi r.{\rm d}r.\Sigma r^2 \Omega$ is the angular momentum 
of an annulus between $r$ and d$r$. Hence the equation (2) multiplied
with $2\pi r.{\rm d}r$ gives how the angular momentum of the annulus changes
by the viscous torque. Therefore, 
\begin{eqnarray}
\Lambda = {1\over{2\pi r}} {{\rm d}G \over{{\rm d} r}},
\end{eqnarray}
where $G(r)$ is the viscous torque at the radial distance $r$.
It is normally assumed that the term ${\rm d}v_{\theta}/{\rm d}r$ 
gives the velocity shear. However for the accretion disk,
\begin{eqnarray} 
{{\rm d}v_{\theta}\over{{\rm d}r}} = {{\rm d}\over{{\rm d}r}}(r\Omega) = 
\Omega + r{{\rm d}\Omega\over{{\rm d}r}},
\end{eqnarray}
in which $\Omega$ is associated with
pure rotation, and $r{\rm d}\Omega/{\rm d}r$ gives the shear.
Therefore, the viscous stress is $\mu r{\rm d}\Omega/{\rm d}r$,
where $\mu$ is the coefficient of viscosity. Then the viscous torque is
\begin{eqnarray}
G(r) = \int r{\rm d}\theta \int{\rm d}z \mu r^2 {{\rm d}\Omega\over{{\rm d}r}}
= 2\pi\nu\Sigma r^3 {{\rm d}\Omega\over{{\rm d}r}}.
\end{eqnarray}
Here, $\nu = \mu/\rho$ is the kinematic viscosity and $\rho$ is the volume density.
Therefore using the equations (2), (3) and (5), we get
\begin{eqnarray}
{\partial\over{\partial t}}(\Sigma r^2 \Omega) + {1\over{r}} {\partial\over{\partial r}}
(\Sigma r^3 \Omega v_r) = {1\over{r}} {\partial\over{\partial r}}
(\nu\Sigma r^3 {{\rm d}\Omega\over{{\rm d}r}}).
\end{eqnarray}
The equations (1) and (6) govern the accretion disk dynamics.
If the disk is nearly Keplerian, then these two equations can be combined to get
\begin{eqnarray}
{\partial\Sigma\over{\partial t}} = {3\over{r}} {\partial\over{\partial r}}
[r^{1/2}{\partial\over{\partial r}}(\nu\Sigma r^{1/2})].
\end{eqnarray}
This equation gives the time evolution of the accretion disk.

Let us now consider a special case, viz., a steady (${\partial\over{\partial t}} = 0$)
geometrically thin accretion disk. From the equations (1) and (6), we get
\begin{eqnarray}
r\Sigma v_r = C_1,
\end{eqnarray}
\begin{eqnarray}
\Sigma r^3 \Omega v_r - \nu\Sigma r^3 {{\rm d}\Omega\over{{\rm d}r}} = C_2,
\end{eqnarray}
where $C_1$ and $C_2$ are constants.
For a steady disk, the mass accretion rate is $\dot{m} = -2\pi r \Sigma v_r$.
Therefore from equation (8),
\begin{eqnarray}
C_1 = -{\dot{m}\over{2\pi}}.
\end{eqnarray}
In order to calculate $C_2$, we consider a boundary condition 
\begin{eqnarray}
{{\rm d}\Omega\over{{\rm d}r}}\arrowvert_{r = r_{\rm in}} = 0.
\end{eqnarray}
Let us justify this condition for the following two cases: (1) when the disk is extended
up to the stellar surface, and (2) when the star is well inside the disk, and the
matter falls from the disk inner edge towards the star quasi-radially.
For a near-Keplerian disk, the angular speed $\Omega$ increases with the
decrease of $r$.
For the first case, since the stellar angular speed is less than Keplerian
(otherwise the star would break up), the disk $\Omega$ must decrease with
the decrease of $r$ close to the star. Therefore, ${{\rm d}\Omega\over{{\rm d}r}}$ must be zero
at some point (assumed to be $r = r_{\rm in}$) near the stellar surface.
On the other hand, for the second case, the torque at the disk inner edge is
expected to vanish (note: it may not be exactly true in reality) making 
${{\rm d}\Omega\over{{\rm d}r}} = 0$. The region (between the star and the disk)
in which the flow deviates considerably from Keplerian is called the 
boundary layer. From the equations (8), (9) and  (11), we get
\begin{eqnarray}
C_2 = -{\dot{m}\over{2\pi}}r_{\rm in}^2\Omega = -{\dot{m}\over{2\pi}}(GMr_{\rm in})^{1/2},
\end{eqnarray}
where $M$ is the stellar mass. Substituting this in the equation (9), and using 
the equations (8) and (10), we get,
\begin{eqnarray}
\nu\Sigma ={\dot{m}\over{3\pi}}[1 - (r_{\rm in}/r)^{1/2}].
\end{eqnarray}
This shows that the accretion rate is greater for larger viscosity.
The viscous dissipation in the accretion disk causes the emission from the
disk, and the source of this emitted energy is the gravitational potential
energy release. This viscous dissipation rate per unit volume of the 
accretion disk is $\mu r^2 ({\rm d}\Omega/{\rm d}r)^2$.
Integrating this over the disk thickness and assuming Keplerian angular
speed, we get the rate of energy emitted per unit area
of the disk from equation (13):
\begin{eqnarray}
-{{\rm d}E\over{{\rm d}t}} = {3GM\dot{m}\over{4\pi r^3}}[1-(r_{\rm in}/r)^{1/2}].
\end{eqnarray}
Therefore, the total power emitted from a disk extending from $r=r_{\rm in}$ to 
infinity is
\begin{eqnarray}
L = {GM\dot{m}\over{2{r_{\rm in}}}}.
\end{eqnarray}
This shows that, for a steady geometrically thin accretion disk in Newtonian gravity, 
half of the gravitational potential energy is emitted by viscous dissipation in the disk,
and the rest half remains stored in the accreted matter as the kinetic energy,
and gets dissipated in the boundary layer. 

%Updated figure from A1744-361 made.
\begin{figure}[btp]
\includegraphics[width=3.00in]{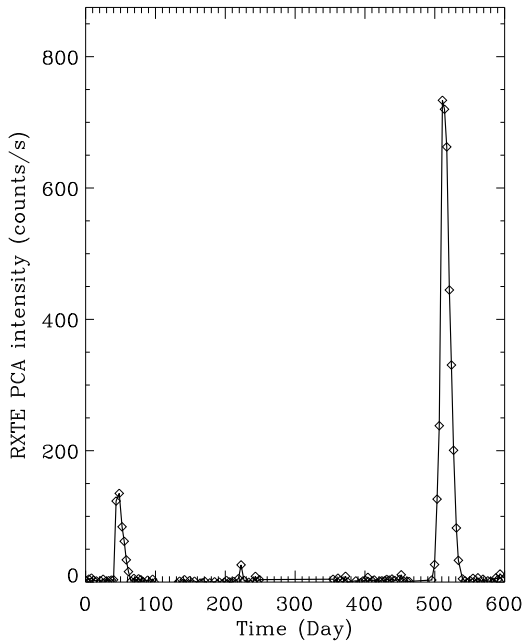}
\caption{Long-term X-ray lightcurve (i.e., observed intensity profile) of a transient 
neutron star LMXB 1A 1744-361 showing its two outbursts \citep{Bhattacharyyaetal2006b}.
The intensity is expressed in the photon count rate detected with the PCA
instrument of the {\it RXTE} satellite.
}
\end{figure}

It is expected that a geometrically thin Keplerian disk should emit like a blackbody.
Therefore, the temperature $T_{\rm eff}(r)$ of the disk surface can be calculated from equation (14):
\begin{eqnarray}
T_{\rm eff}(r) = \{{3GM\dot{m}\over{8\pi r^3}\sigma}[1-(r_{\rm in}/r)^{1/2}]\}^{1/4},
\end{eqnarray}
where $\sigma$ is the Stefan-Boltzmann constant. As the blackbody temperature
varies with the radial distance, the spectrum from a disk is known as
multicolor blackbody. For luminous LMXBs, the X-rays from the neutron star
boundary layer and the inner part of the accretion disk can substantially
irradiate the disk surface. This irradiation temperature $T_{\rm irr}(r)$
(which is blackbody temperature) is proportional to $r^{-1/2}$ \citep{Kingetal1996}.
Therefore, the net effective disk temperature is $T_{\rm disk}(r) =
[T^4_{\rm eff}(r) + T^4_{\rm irr}(r)]^{1/4}$, in which
$T_{\rm eff}(r)$ dominates for the inner portion (that emits X-rays) 
of the disk, and $T_{\rm irr}(r)$ dominates far away from the neutron star. Note that
the latter can have significant consequences for the disk instability \citep{Kingetal1996}.

\section{Selected Features}

\subsection{Outbursts}

A class of neutron star LMXBs are called transients.
These sources have two distinctly different intensity states: (1) the
quiescent state in which the source remains for months to years; and
(2) the outburst state which continues for weeks to months (sometimes
for years; e.g., for the source KS 1731-260). 
The luminosity of the source typically increases by
several orders of magnitude as it evolves from the quiescent state 
into an outburst. Fig. 3 shows the long term intensity profile of such a 
transient 1A 1744-361, exhibiting two clear outbursts. 
The transient nature of these LMXBs is normally attributed to an
accretion disk instability, that causes high accretion rates for
certain periods of time, and almost no accretion during other times
(see King 2001 \citep{King2001} for a review). These sources are
often not detectable in their quiescent states, and in some
cases when they are detectable, the poor statistics due to the
small number of observed photons hinders a detailed study. 
For most of the scientific purposes, it is therefore preferable to
observe these transients during their outbursts. New transients
are also typically discovered during the outburst states.
However, it cannot normally be predicted when a transient will
come out of its quiescent state. Therefore, instruments with
large field of view that can monitor the entire sky almost 
continuously are required to find the transients in their
outbursts. The {\it RXTE} ASM is such a very useful
instrument, that has discovered many new transient LMXBs, as well
as detected subsequent outbursts of them.

However, there are scientific reasons that motivate the observations
of transient neutron star LMXBs in their quiescent states. For example,
during the quiescence, the neutron star surface should be the primary X-ray 
emitting component, which can be modelled to constrain the stellar radius.
Besides, a long outburst heats up the neutron star crust taking it out of 
thermal equilibrium with the rest of the star. Therefore, tracking the
LMXB system going into the quiescence, and the analysis of the quiescent X-ray data
can provide a unique opportunity to probe some aspects of the neutron star 
extreme physics (e.g., Cackett et al. \citep{Cackettetal2008b}). 
For example, such a study allows one to compare the observed neutron star cooling rate
with the theoretical cooling curve, that involves the physics of stellar constituents.
Note that the satellites (e.g., {\it Chandra}, {\it XMM-Newton}) with high sensitivity
imaging instruments are used to observe the sources in quiescence.

%Figure from 4U 1636-536 made.
\begin{figure}[btp]
\includegraphics[width=3.00in]{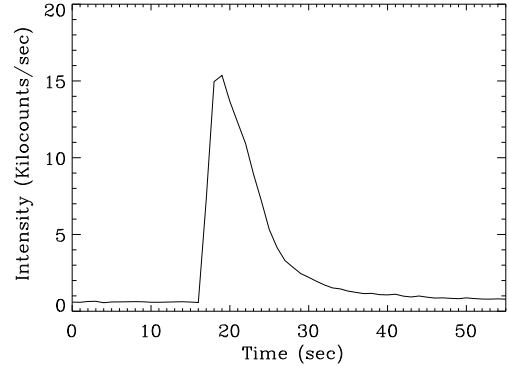}
\caption{X-ray intensity profile of a typical thermonuclear X-ray burst
observed from an LMXB. After the onset of the burst, the X-ray intensity
increases quickly by more than an order of magnitude (compared to the non-burst
intensity corresponding to the gravitational energy release), and then 
decays relatively slowly as the neutron star surface cools down.
}
\end{figure}

\subsection{Thermonuclear X-ray Bursts}

Many neutron star LMXB systems show a spectacular eruption in X-rays
every few hours to days. During these bursts, the observed X-ray intensity
goes up sharply in $\approx 0.5-5$ seconds, and decays relatively slowly in 
$\approx 10-100$ seconds (see Fig. 4; Strohmayer \& Bildsten \citep{StrohmayerBildsten2006}). 
The typical energy emitted in a few seconds during such a burst is $\sim
10^{39}$ ergs (note that the Sun takes more than a week to release this
energy). These bursts are known as type-I or thermonuclear bursts, and 
were first discovered in 1970's \citep{Grindlayetal1976,
Belianetal1976}. Soon after their discovery it was realized that they
originate from recurrent unstable nuclear burning of accreted matter 
on the neutron star surfaces \citep{Joss1977, LambLamb1978, StrohmayerBildsten2006}. 
The neutron star surface origin was supported by the observational
fact that the burst emission area (estimated from the energy spectrum
of the bursts) matched with the expected surface area 
of a neutron star \citep{Swanketal1977}. Although, a competetive emission area could
be the inner portion of the accretion disk, nuclear ignition in order to
produce the observed bursts cannot occur in such a rare medium.
But what shows that the nuclear burning causes the bursts?
Before coming to that point, we would like to convince the readers that
such a nuclear burning has to be unstable. This is because, while the
nuclear energy released per nucleon by fusion is only a few MeV, the
gravitational energy release by a nucleon falling on the neutron star surface
from a large distance is $GMm_{\rm nucleon}/R$ ($M$ and $R$ are neutron star mass 
and radius respectively), which is typically $\sim 200$ MeV. Therefore, 
the energy released from a stable burning of the accreted matter would be 
entirely lost in the observed gravitational energy. So the nuclear burning
can power the bursts, only if the accreted matter accumulates on the neutron
star surface for a long time, and then burns quickly in an unstable manner.
This instability appears if the nuclear energy generation rate is more temperature 
sensitive than the radiative cooling rate.
For an observed data set from a given source,
the ratio of the total energy content in all the bursts to that in the non-burst
emission roughly tallies with the expected nuclear energy release (per nucleon)
to the expected gravitational energy release (per nucleon) ratio \citep{Bildsten2000}
This is one of the strongest evidences of the nuclear energy origin of the bursts. 
Here we note that,
only a year before these bursts were discovered, Hansen and Van Horn \citep{HansenvanHorn1975}
predicted that thin hydrogen and helium layers on neutron star surfaces could be
susceptible to the above mentioned instability \citep{StrohmayerBildsten2006}.

The accreted material consists of mostly hydrogen, some helium and a small amount
of heavier elements, except when the donor companion is a white dwarf.
This matter, that accumulates on the neutron star surface, undergoes 
hydrostatic compression as more and more material piles on, and the ignition density
and temperature are reached in a few hours to days. 
The burning typically happens at a depth of $\approx 10$ m at a column density 
of $\sim 10^8$ gm cm$^{-2}$. The compression rate
and the ignition depends on the accretion rate, which sets four distinct 
regimes of burning \citep{Bildsten2000}. 
{\it Regime 1}: in most cases, the temperature of the accumulated material 
exceeds $10^7$ K, and hence the hydrogen 
burns via the CNO cycle instead of the proton-proton (pp) cycle.
This causes mixed hydrogen and helium bursts triggered by the unstable
hydrogen ignition, which happens for the accretion rate per unit neutron star 
surface area \.{m} $ < 900$ gm cm$^{-2}$ sec$^{-1}$ 
($Z_{\rm CNO}/0.01$)$^{1/2}$ \citep{Bildsten2000}.
Here $Z_{\rm CNO}$ is the mass fraction of CNO. 
{\it Regime 2}: at higher
temperatures ($T > 8\times10^7$ K), the proton capture timescale becomes
shorter than the subsequent $\beta$ decay lifetimes. Hence, the hydrogen
burns via the ``hot" CNO cycle of Fowler \& Hoyle \citep{FowlerHoyle1965}: 
$$^{12}{\rm C}(p,\gamma)^{13}{\rm N}(p,\gamma)^{14}{\rm O}(\beta^{+})^{14}{\rm N}(p,\gamma)^{15}{\rm O}(\beta^{+})^{15}{\rm N}(p,\alpha)^{12}{\rm C}.$$
In this case, the hydrogen burning happens 
in a thermally stable manner (i.e., without triggering
a thermonuclear burst) simultaneously with the accumulation of matter. 
This allows a helium layer to build up below the hydrogen layer. For the
\.{m} range $900$ gm cm$^{-2}$ sec$^{-1}$ ($Z_{\rm CNO}/0.01$)$^{1/2}
< $ \.{m} $< 2\times10^3$ gm cm$^{-2}$ sec$^{-1}$ ($Z_{\rm CNO}/0.01$)$^{13/18}$,
the hydrogen becomes entirely burned before this helium is ignited \citep{Bildsten2000, 
StrohmayerBildsten2006}. So when
the helium ignition happens, a short ($\sim 10$ sec) but very intense burst occurs by the
unstable triple-alpha reaction of the pure helium ($3\alpha$ $\rightarrow$ $^{12}{\rm C}$).
Such a burst is called a ``helium burst". Many of these bursts are so intense,
that the local X-ray luminosity in the neutron star atmosphere may exceed the
Eddington limit (for which the radiative pressure force balances the gravitational 
force), and the photospheric layers may be lifted off the neutron star surface.
Such bursts are called photospheric radius expansion (PRE) bursts.
{\it Regime 3}: for \.{m} $> 2\times10^3$ gm cm$^{-2}$ sec$^{-1}$ 
($Z_{\rm CNO}/0.01$)$^{13/18}$, hydrogen burns via the ``hot" CNO cycle in a 
stable manner (as for the regime 2), but enough unburnt hydrogen remains
present at the time of helium ignition. This is because at this higher 
accretion rate, the helium ignition conditions are satisfied much sooner.
Therefore in this regime, mixed hydrogen and helium bursts are triggered
by the helium ignition. During such a burst, the thermal instability
can produce elements beyond the iron group \citep{Hanawaetal1983, HanawaFujimoto1984, 
Koikeetal1999, Schatzetal1999, Schatzetal2001} via the rapid-proton (rp) process of Wallace \&
Woosley \citep{WallaceWoosley1981}. This process burns hydrogen via successive proton
capture and $\beta$ decays. The long series of $\beta$ decays makes these
bursts typically much longer ($\sim 100$ sec) than the helium bursts.
{\it Regime 4}: finally at a very high accretion rate (comparable to the Eddington limit),
the helium burning temperature sensitivity becomes weaker than the cooling 
rate's sensitivity \citep{AyasliJoss1982, Taametal1996}. Therefore in this
regime, the stable burning sets in, and the thermonuclear bursts do not occur.

Apart from type-I X-ray bursts, another type of thermonuclear X-ray bursts
with much larger recurrence time ($\sim$ years) has been observed.
These are called superbursts, and they differ from the type-I bursts also
in the decay time ($\sim 1-3$ hours) and the amount of released energy
($\sim 10^{42}$ ergs). The first superburst was discovered by Cornelisse
et al. \citep{Cornelisseetal2000} from the known type-I burster 4U 1735-44. So far about 15 
superbursts have been detected from 10 sources \citep{Cooperetal2009}.
It is believed that these bursts of large fluence happen by the $^{12}C+^{12}C$ 
fusion reaction at a column depth of $\approx 10^{12}$ gm cm$^{-2}$ 
 \citep{CummingBildsten2001, StrohmayerBrown2002, Cooperetal2009}.
However, currently our understanding of superbursts is not as clear as that
of type-I bursts.

Thermonuclear bursts provide a unique opportunity to study some aspects of
extreme physics. Moreover, such a study brings together several branches 
of physics: nuclear physics, astrophysics, general theory of relativity,
magnetohydrodynamics, etc. These bursts can also be useful to constrain the 
neutron star parameters, and hence to understand the nature of the stellar
core matter at supranuclear densities (see the spectral and timing properties
of these bursts in later sections). Therefore, the thermonuclear bursts
belong to a truly multidisciplinary field, which is not yet well understood.
One such poorly explored aspect of these bursts is the thermonuclear
flame spreading on the neutron star surfaces. Burst is ignited
at a certain location on the stellar surface, and then the burning region 
expands to cover the entire surface \citep{FryxellWoosley1982, 
Spitkovskyetal2002}. The theoretical modeling of thermonuclear flame spreading
including all the main physical effects is extremely difficult, and only
recently Spitkovsky et al. \citep{Spitkovskyetal2002} have provided some
insights in this field. Although these authors have ignored several physical
effects (magnetic field, strong gravity, etc.) in their model, they have
considered the effects of the Coriolis force, which is important as the
bursting neutron stars are rapidly spinning (spin frequency $\sim 300-600$ Hz).
The observational study of flame spreading is also no less difficult than 
the theoretical study. This is because, this spreading mostly happens
within $\approx 1$ sec during the burst rise, and in order to study it
observationally, one needs to detect and measure the evolution of the burst
spectral and timing properties during this short time. Only recently
this has been possible for a few bursts by analyzing the {\it RXTE} PCA data
 \citep{BhattacharyyaStrohmayer2006a, BhattacharyyaStrohmayer2006b, 
BhattacharyyaStrohmayer2007a}. Moreover, an indication of
the effects of the Coriolis force on the flame spreading (as predicted by
Spitkovsky et al. \citep{Spitkovskyetal2002}) 
has also been found \citep{BhattacharyyaStrohmayer2007c}.

\subsection{Type-II Bursts}

%Updated figure for A1744 made.
\begin{figure}
\includegraphics[width=3.00in]{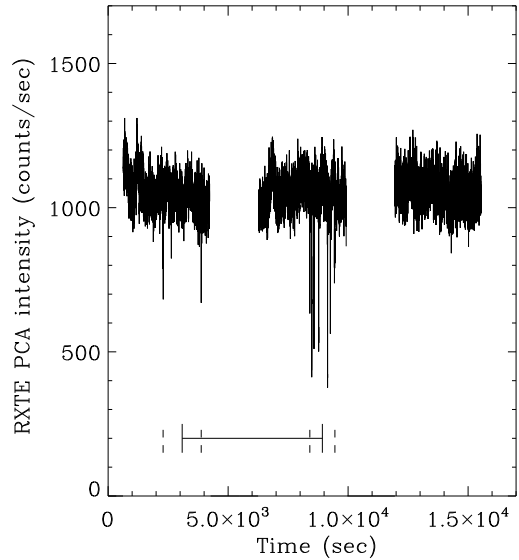}
\caption{X-ray intensity dips observed from a neutron star LMXB 1A 1744-361
 \citep{Bhattacharyyaetal2006b}. The separation between two sets of consecutive dips
gives the orbital period of the binary system (see text). 
Because of the gap in the observed data, one cannot be sure whether 
the two sets of dips shown in this figure are consecutive. Hence,
these two sets of dips may be the orbital period or a multiple of that.
}
\end{figure}

A series of bursts, known as type-II X-ray bursts, are
observed from neutron star LMXBs in quick succession.
To the best of our knowledge, only two such systems (rapid burster and GRO J1744-28)
exhibit these bursts \citep{Lewinetal1976, Fishmanetal1995, Lewinetal1996}.
The intervals between the bursts vary between seconds to $\sim 1$ hour,
and the burst duration is in the range of seconds to minutes. The typical
emitted energy range during a burst from the rapid burster is
$\sim 1\times10^{38}$ ergs to $\sim 7\times10^{40}$ ergs \citep{Lewinetal1995}. 
These bursts often show successive peaks and flat tops. Note that
a source is not always active for type-II X-ray bursts, and a series
of such bursts are observed in certain periods of time. type-II X-ray bursts
are distinctly different from the type-I bursts in many ways. For example,
The latter bursts show a spectral softening during burst decay, while
the former bursts lack this signature. type-II bursts are believed to
be caused by spasmodic accretion, possibly due to an accretion instability \citep{Lewinetal1995}.
Therefore, these bursts are powered by the gravitational energy,
while the type-I bursts are powered by the nuclear energy.
The recent models of the type-II bursts include the effects of
magnetosphere \citep{SpruitTaam1993} and intermittent advection-dominated
accretion flow \citep{Mahasenaetal2003}. However, the origin of the
accretion instability is not yet well understood.

%No figure.

\subsection{Dips and Eclipses}

Some neutron star LMXBs show periodic temporary decrease in X-ray intensity.
This timing feature can be divided into two distinct categories: eclipses
and dips. Both of them occur with the binary orbital period, and hence 
provide an excellent way to measure this period \citep{WhiteMason1985}.
Partial and total eclipses are believed to be caused by the obscuration
of the X-ray source by the companion star \citep{Parmaretal1986}. 
Such an obscuration requires an observer's line-of-sight close to
the plane of the binary system (i.e., close to the accretion disk plane).
Therefore, the observation of eclipses from a source implies that its
inclination angle is large ($> 75^{\rm o}$; Frank et al. \citep{Franketal1987}). 
This inclination angle $i$ is measured from the direction perpendicular to
the accretion disk plane. The dips are periodic but irregular, and 
typically are more pronounced at lower energies \citep{Parmaretal1986}.
The shape, depth and duration of each dipping interval widely vary from 
cycle to cycle, and within such an interval the observed intensity varies
dramatically on a timescale  of a few seconds (see Fig. 5). These
dips are believed to be caused by a bulge on the edge of the accretion
disk at the region where the gas stream from the companion star hits the
disk (see Fig. 2 for a view of this impact region; White \& Swank \citep{WhiteSwank1982}).
Therefore, dips imply $i > 60^{\rm o}$ \citep{Franketal1987}.
The detailed nature of this bulge is still a matter of debate, and
spectral analysis can shed light on this. 

%Updated version of disk line from Ser X-1.
\begin{figure}
\includegraphics[width=3.00in]{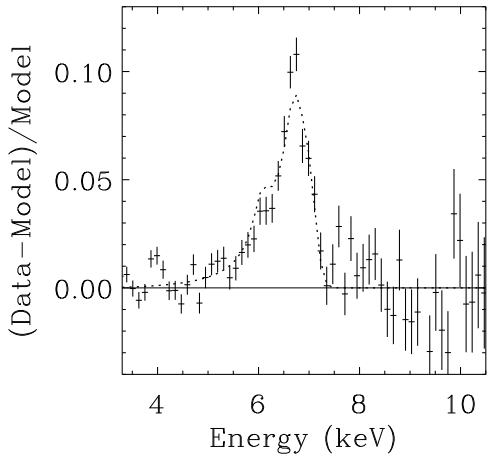}
\caption{Broad relativistic iron (Fe K$\alpha$) spectral line from 
the inner accretion disk of a neutron star LMXB Serpens X-1
 \citep{BhattacharyyaStrohmayer2007b}. The x-axis shows the X-ray energy,
and the y-axis gives the observed intensity in excess to the best-fit
continuum spectral model. The data points (shown with error bars) clearly
shows an asymmetric spectral line, while the dotted profile is a 
relativistic model of spectral line that fits the observed line well.
}
\end{figure}

%Figure got from Mariano.
\begin{figure*}[h]
\includegraphics[width=6.00in]{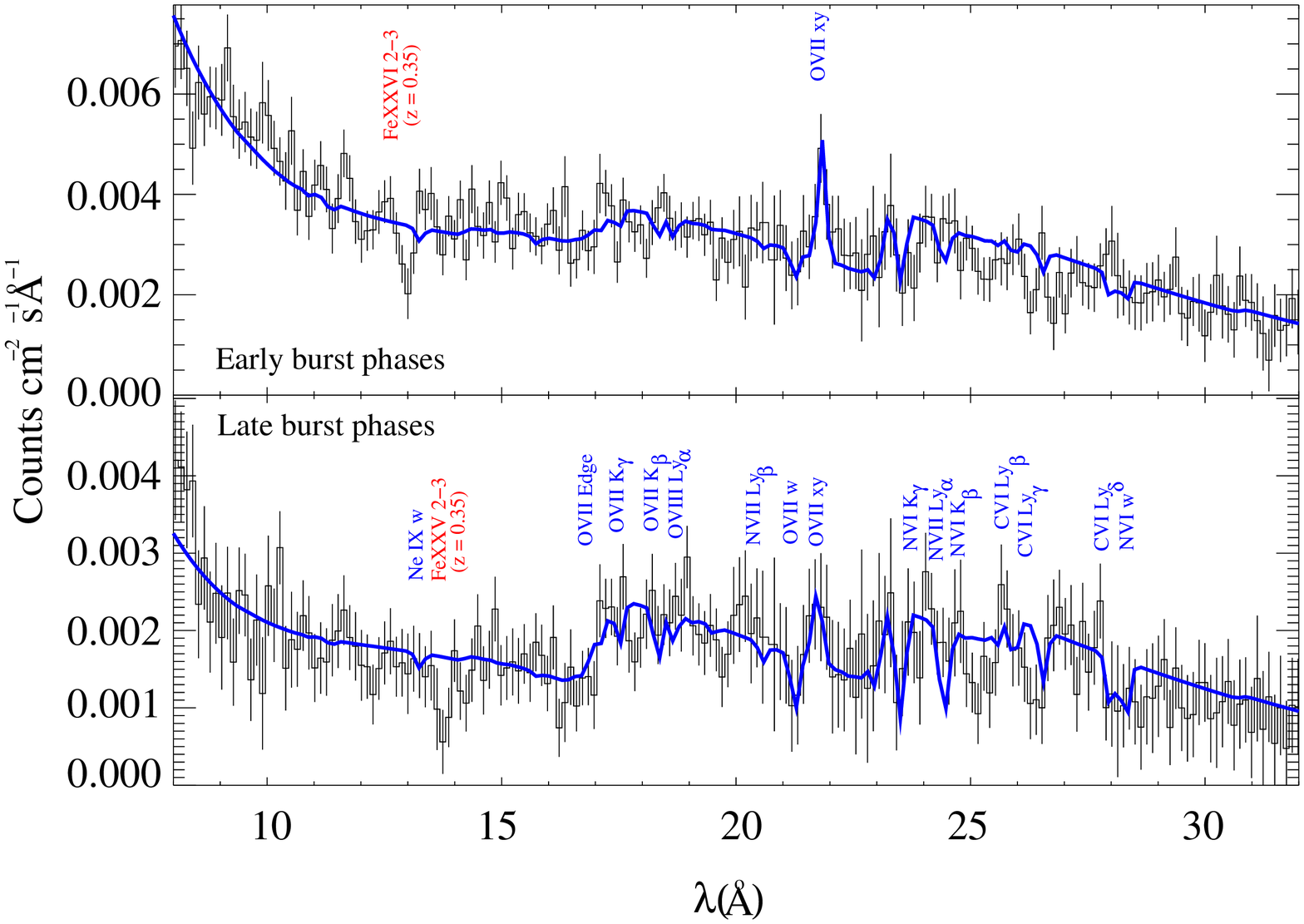}
\caption{The total background-subtracted {\it XMM-Newton} RGS energy spectra 
(intensity vs. wavelength) 
of the neutron star LMXB EXO 0748-676 during 28 thermonuclear X-ray bursts.
The data are plotted as the black histograms with $1\sigma$ error bars.
The blue line is the empirical continuum, with additional O VII intercombination
line emission, modulated by absorption in photoionized circumstellar material.
The upper panel is for the early burst phases (when the burst intensity is more),
while the lower panel is for the late burst phases (when the burst intensity is less).
The two red labels indicate the iron absorption lines at a
gravitational redshift $z=0.35$ plausibly originated from the neutron star 
atmosphere (figure courtesy: Mariano M\'endez; Cottam et al. \citep{Cottametal2002}).
}
\end{figure*}

\section{Spectral Aspects}

\subsection{Continuum Spectrum}

\subsubsection{Persistent Emission}

The study of the continuum spectrum of the persistent emission 
(i.e., non-burst, non-dipping, non-eclipsing) from the neutron star 
low-mass X-ray binary systems can be very useful to understand 
the various components of the accretion flow and their structures,
including the flow in the strong gravity region near the neutron star.
Such a spectrum can be typically fitted with two spectral components.
Mitsuda et al. \citep{Mitsudaetal1984} used a single-temperature blackbody plus a multicolor
blackbody model. The former component is expected to be produced from
a boundary layer between the accretion disk and the neutron star, while
the latter may be produced from the accretion disk. In an alternative
two-component model, White et al. \citep{Whiteetal1985, Whiteetal1986}
used a Comptonization component 
instead of a multicolor blackbody. They assumed that the Comptonization 
component was originally the disk component which was dominated by the 
Comptonization processes. In the later years, it has been found that
a Comptonization component is often required to fit the observed 
persistent emission continuum spectrum, and an additional
blackbody is needed when the source is more luminous and spectrally softer
(see, for example, Bhattacharyya et al. \citep{Bhattacharyyaetal2006c}). 
Apart from these, there is evidence of an extended ``accretion disk 
corona (ADC)" for some sources, that can be modelled with a power law \citep{Churchetal1997}.
However, we note that
much work needs to be done in order to build a self-consistent spectral
model involving all the X-ray emitting and absorbing components, and 
to fit the observed spectrum with such a single model. 

%No figure.

\subsubsection{Thermonuclear Burst Emission}

Spectrum of a thermonuclear X-ray burst is obtained usually by subtracting
the preburst emission spectrum (assumed to be powered by the gravitational
energy release) from the observed spectrum during the burst \citep{Gallowayetal2008}.
The continuum burst spectra are thermal, and in most
cases they are quite well described with a blackbody model \citep{Swanketal1977,
Hoffmanetal1977}. The temperature of this blackbody naturally
increases rapidly during the burst intensity rise, and then decreases relatively
slowly during the burst decay. If the entire neutron star surface emits
like a blackbody of the same temperature at a given time, then the neutron star
radius ($R_{\rm obs}$) can, in principle, be measured from the following
formula: $R_{\rm obs} = d[F_{\rm obs}/(\sigma T^4_{\rm obs})]^{1/2}$.
Here, $F_{\rm obs}$ and $T_{\rm obs}$ are the observed flux and
temperature respectively, and $d$ is the source distance, which may be
known by an independent technique. Note that the measurement of the neutron
star radius can be useful to understand the nature of the dense matter
of the stellar core. However, in order to do such measurement, one needs
to include the following two effects. (1) The electron (Compton) scattering
in the neutron star atmosphere shifts the blackbody spectrum (originated from 
the burning layer) towards the higher energies, and changes its shape
slightly \citep{Londonetal1984, Londonetal1986, EbisuzakiNakamura1988, Madej1991,
Titarchuk1994}. Such a blackbody is called a ``diluted blackbody", and its
temperature ($T_{\rm col}$, i.e., the color temperature) is related with
the blackbody temperature ($T_{\rm BB}$) by the color factor $f$.
(2) The surface gravitational redshift shifts the temperature towards a 
lower value. Therefore, the observed temperature $T_{\rm obs}$ is related
to $T_{\rm BB}$ by $T_{\rm BB} = T_{\rm obs} (1+z)/f$. Correspondingly, the
actual neutron star radius $R_{\rm BB}$ is then related to the observed
radius $R_{\rm obs}$ by $R_{\rm BB} = R_{\rm obs} f^2/(1+z)$. Therefore,
in order to measure the neutron star radius from the continuum burst spectra,
one needs to independently estimate or measure the color factor $f$ and the surface
gravitational redshift factor $1+z$. The former can be estimated from the 
temperature, surface gravity and the atmospheric chemical 
composition \citep{Madejetal2004}, while the latter can be calculated from the neutron
star radius-to-mass ratio. This ratio can be independently measured from
the surface atomic spectral lines or the burst oscillations (see later).

The evolution of the continuum spectrum of a thermonuclear burst can be
best measured using the PCA instrument of the {\it RXTE} satellite.
Such a measurement during the burst rise will be extremely useful to
study the thermonuclear flame spreading. The chemical composition of
the neutron star atmosphere may change during the burst. Therefore,
and since the burst temperature also evolves, the observed radius $R_{\rm obs}$
changes substantially, and apparently irregularly during the burst
decay \citep{Gallowayetal2008}. An understanding of these changes will be
essential to (1) measure the neutron star radius using the burst continuum
spectrum, (2) understand the nuclear physics of bursts, and (3) probe
the magnetohydrodynamics of the burning layer and the atmosphere.

%No figure.

\subsubsection{Dipping Emission}

In order to probe the nature of the bulge that causes dips, it is essential to
understand how the continuum spectrum changes as the source goes into the dipping interval.
A simple increase in the neutral absorber column density from its persistent emission
value cannot explain the dip spectra. Currently, there are two competing spectral 
models to explain dips. (1) The ``progressive covering" model (e.g., Church et 
al. \citep{Churchetal1997}): 
in this approach, the X-ray emission is considered to originate from a 
point-like blackbody and a power-law from an extended component (e.g., a corona). 
The dipping is assumed to be caused by the partial and progressive covering of 
the extended component by a neutral absorber.
(2) The ionized absorber model (e.g., Boirin et al. \citep{Boirinetal2005}):
in this approach, the dips are caused by the increase in column density
and the decrease in ionization state of an ionized absorber, along with
a neutral absorber column density increase. This model can naturally explain some of the
observed narrow spectral features.

\subsection{Line Spectrum}

\subsubsection{Lines from High Inclination Sources}

%No figure.

Many of the high inclination neutron star LMXBs (for example, dipping sources)
exhibit narrow spectral absorption features (e.g., Boirin et 
al. \citep{Boirinetal2004}; Parmar et al. \citep{Parmaretal2002}; 
D\'iaz Trigo et al. \citep{Trigoetal2006}), 
as well as narrow recombination emission lines
 \citep{JimenezGarateetal2003}. The most prominent of these lines are Fe XXV
and XXVI absorption lines. This spectral features are believed to originate
from the material above the accretion disk, and can be observed only when the
observer's line-of-sight passes through this material. This is possible only
for the high inclination sources. The narrow spectral features, therefore,
can be very useful to constrain the structures and properties (e.g., ionization
states, relative abundances of various elements) of the photoionized plasma 
above the accretion disks in these systems \citep{JimenezGarateetal2003}. 
Note that, although the
spectral features are mostly seen from high inclination LMXBs, this plasma
must be common to all LMXBs, and hence these
features will be useful in understanding LMXBs in general.

\subsubsection{Line from Disk}

Broad asymmetric spectral iron emission line has been observed from many
accreting supermassive and stellar-mass black hole systems \citep{ReynoldsNowak2003,
Miller2006}. It is believed to originate from the inner portion of the 
accretion disk, when the disk is irradiated by a hard X-ray source (e.g., an
accretion disk corona). Such an irradiation creates a ``reflection" component
of the X-ray spectrum. When the hard X-ray photon enters the colder disk,
it is either scattered out of the disk, or destroyed by Auger de-excitation,
or reprocessed into a fluorescent line photon that escapes the 
disk \citep{Fabianetal2000}. 
The observed iron K$\alpha$ line near 6.4 keV is the strongest 
fluorescent line from such a system.
As the matter in the inner accretion disk rotates very fast, and is close to the black hole,
Doppler effect broadens the line, special relativistic beaming makes it asymmetric,
and the gravitational redshift shifts it towards the lower energies. Therefore,
these lines can be useful to study the accretion flow close to the black hole, 
to test the theory of general relativity, and to measure the black hole spin.

For long time it has been known that the persistent spectra of many
neutron star LMXBs contain broad iron emission line \citep{Asaietal2000}.
It was suspected that such a line might have an origin similar to that
of the broad iron line from black hole systems.
However, the asymmetry of these lines was not established, and hence 
their origin was not well understood. 
Recently, Bhattacharyya \& Strohmayer \citep{BhattacharyyaStrohmayer2007b}
have, for the first time, established the inner disk origin of such
a line from the neutron star LMXB Serpens X-1 (see Fig. 6). This was later supported 
for several other sources (e.g., Cackett et al. \citep{Cackettetal2008a}; 
Pandel et al. \citep{Pandeletal2008};
Di Salvo et al. \citep{DiSalvoetal2009}). Therefore, the high precision observations 
of these broad iron K$\alpha$ line
from neutron star LMXBs could be useful to constrain the stellar
parameters (radius, angular momentum) and to probe the inner accretion disk.
However, the evidence of the ``reflection" component of the X-ray spectrum
is still marginal \citep{DiSalvoetal2009}, which the theory and the future
observations should address.

\subsubsection{Line from Neutron Star}

Observing atomic spectral lines from the surface of a neutron star has been one of the 
long-standing goals of the neutron star astrophysics. This is because, such an 
observation and the identification of the line provide the cleanest way to measure the 
neutron star radius-to-mass ratio using the following formula (for the Schwartzschild
spacetime appropriate for a non-spinning neutron star): 
$E_{\rm em}/E_{\rm obs} = 1+z = [1-(2GM/Rc^2)]^{-1/2}$ \citep{OzelPsaltis2003, 
Bhattacharyyaetal2006a}. Here, $E_{\rm em}$ is the emitted energy
of the line photons, $E_{\rm obs}$ is the observed energy of these photons (different
from  $E_{\rm em}$ because of the surface gravitational redshift), and $M$ and $R$ are the mass and the
radius of the neutron star respectively. 
Note that the measurement of the stellar parameter $R/M$ can be useful to 
understand the nature of the dense core matter of neutron stars.
As the neutron stars in LMXB systems typically
spin with high frequencies ($\sim 300-600$ Hz), Doppler effect and special relativistic 
beaming should make the lines broad and asymmetric (similar to the lines from disks
mentioned in the previous section). Measuring $R/M$ accurately from such a broad line
(for which the definition of $E_{\rm obs}$ is not certain) using the above formula is difficult.
Bhattacharyya et al. \citep{Bhattacharyyaetal2006a} have recently suggested 
a way to measure the stellar $R/M$ using the above formula from such
lines. These broad asymmetric 
lines can have low-energy and high-energy peaks (or dips for absorption
lines). Bhattacharyya et al. \citep{Bhattacharyyaetal2006a} have found that the ratio of energy contents
of these peaks (or, dips) contains information about the general relativistic frame-dragging.
Therefore, surface atomic lines can be used to test the theory of general
relativity in the strong gravity regime.

Surface lines are likely to be detected from thermonuclear X-ray bursts and hence from
neutron stars in LMXB systems. This is because: (1) the continuous accretion and
the radiative pressure may keep the line-forming heavy elements in the stellar atmosphere
for the time-duration required for the line detection; (2) the relatively low magnetic fields
($\sim 10^7-10^9$ Gauss) 
of the neutron stars in LMXBs should make the line identification easier by keeping the 
magnetic splitting negligible; (3) the lower magnetic field does not complicate the
modeling of the atmosphere; (4) the luminous bursts give good signal-to-noise ratio;
and (5) during the bursts, typically 90\% of the total emission originate from the
neutron star surface, and hence the surface spectral line may be largely free from the 
uncertainty due to the other X-ray emission components, such as the accretion disk.

Indeed, the only plausible observation of the neutron star surface atomic spectral
lines was from an LMXB EXO 0748-676 \citep{Cottametal2002}. These authors detected
two significant absorption features in the {\it XMM-Newton} RGS energy spectra of the 
thermonuclear X-ray bursts, which they identified as 
Fe XXVI and Fe XXV $n=2-3$ lines with a redshift of $z = 0.35$ (see Fig. 7).
In order to detect these lines, they had to combine the spectra of 28 bursts.
The identification of these two lines as the surface atomic spectral lines
is reasonable for the following reasons:
(1) the early burst phases (when the temperature was higher) showed the Fe XXVI
line, while the late burst phases (when the temperature was lower) showed the Fe XXV
line. This is consistent with qualitative temperature dependence of the
iron ionization states. (2) The narrow lines were consistent with the relatively slow
spin rate of the neutron star (45 Hz; Villarreal \& 
Strohmayer \citep{VillarrealStrohmayer2004}).
(3) Both the lines showed the same surface gravitational redshift, as expected.
This redshift is reasonable for the surface of a typical neutron star.
The absorption of some of the bursts photons (originated in deep burning layers) 
by the iron ions in the upper atmosphere might give rise to these redshifted
lines. Chang et al. \citep{Changetal2005} showed that the observed strength of the iron lines
could be produced by a neutron star photospheric metallicity, which was 
$2-3$ times larger than the solar metallicity.
However, these lines were not significantly detected from the new data from the
later observations of the same source. Although, this casts doubts in the reality
of these lines, it may be caused by the change in the photospheric conditions 
that might weaken the lines.
Therefore, the evidence of the existence of surface atomic lines currently
remains uncertain, which, given the importance of these lines, provides the 
motivation for the future generation X-ray instruments. 

\section{Timing Aspects}
 
\subsection{Color-color diagram}

%Figure from a Z source given by Jonker.
\begin{figure}
\includegraphics[width=3.00in]{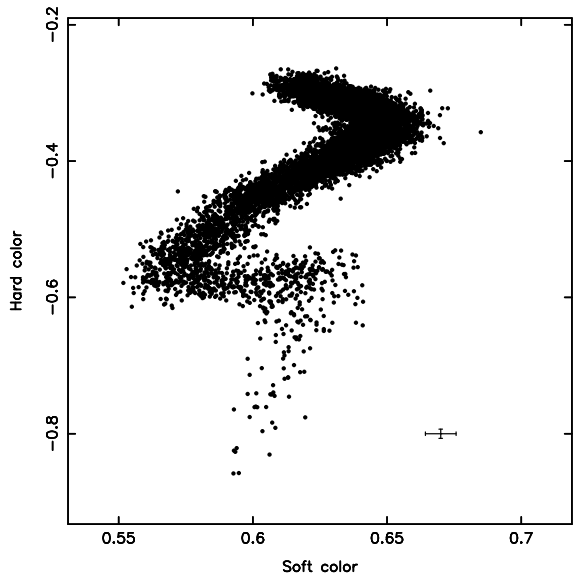}
\caption{Typical color-color diagram (see text for the definition)
of the members of the ``Z'' subdivision of neutron star LMXBs.
Such a diagram shows the evolution of the source spectral state
(figure courtesy: Peter Jonker; Jonker et al. \citep{Jonkeretal1998}).
}
\end{figure}

%Figure from an atoll source given by Diego.
\begin{figure}
\includegraphics[width=3.00in]{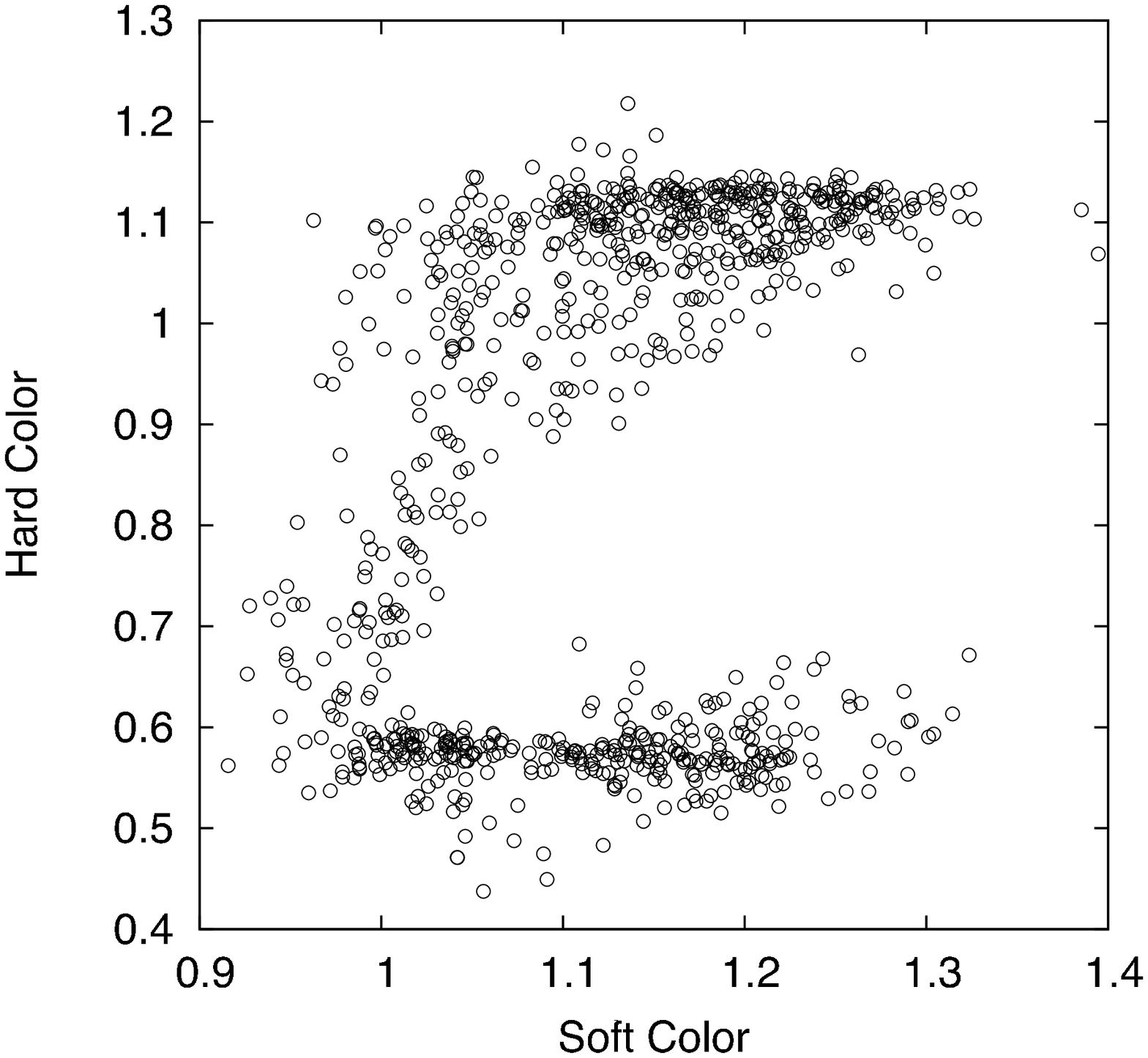}
\caption{Typical color-color diagram (see text for the definition)
of the members of the ``atoll'' subdivision of neutron star LMXBs
(figure courtesy: Diego Altamirano; van Straaten et al. \citep{vanStraatenetal2003}; 
Altamirano et al. \citep{Altamiranoetal2008b}).
}
\end{figure}

A convenient way to study how the persistent spectrum of a neutron star LMXB changes
with time, and how it is correlated with various timing properties and the
occurrence of other features (e.g., thermonuclear X-ray bursts), is to compute
the color-color diagram. In order to do this, typically five photon energies
with increasing values are considered: $h\nu_1$, $h\nu_2$, $h\nu_3$, $h\nu_4$ and
$h\nu_5$. Let us assume that the observed background-subtracted photon count
numbers in the ranges $h\nu_1-h\nu_2$, $h\nu_2-h\nu_3$, $h\nu_3-h\nu_4$ and $h\nu_4-h\nu_5$
are $N_1$, $N_2$, $N_3$ and $N_4$ respectively. Then we define soft color $=
N_2/N_1$ and hard color $= N_4/N_3$. The color-color diagram is a plot of hard color
vs. soft color, which shows the change of spectral hardness of the source.
Neutron star LMXBs can be divided into two classes based on their color-color diagrams:
the so called ``Z sources", that show distinct Z-like curves (Fig. 8), and the
``atoll sources", that exhibit C-like curves (Fig. 9). These two types
of curves are not only different by shape, but also they are correlated with  
distinctly different timing properties \citep{vanderKlis2006}. Apart from these differences,
the Z sources are very luminous (luminosities are close to Eddington), while
the luminosities of the atoll sources typically remain between $0.01-0.2$ of the
Eddington luminosity \citep{vanderKlis2006}. The three branches of a Z curve,
from upper to lower, are called horizontal branch (HB), normal branch (NB) and
flaring branch (FB). The lower part of a typical atoll curve is called the
banana state, while the upper part is known as the island state. The X-ray 
luminosity typically increases from HB to FB for a Z source, and from
island state to banana state for an atoll source. We note that, although
the various characteristics of the color-color diagrams are observationally
very robust, most of them are not yet theoretically well understood. Accretion
rate is known to be a parameter that changes along the Z curves and the atoll
curves. But, given the complex properties of these curves, more parameters
must be involved in the fundamental level. It is also not understood what,
apart from the accretion rate, distinguishes between the Z and atoll properties.
The recent discovery of the first transient Z source, that shows substantially
changed properties at lower luminosities, holds the promise to take our 
understanding of these sources to a higher level \citep{Homanetal2007}.

\subsection{Regular Pulsations}

Most of the radio pulsars (that are spinning neutron stars emitting periodic 
radio pulses) have spin period $> 10$ milliseconds, and high magnetic field
(typically $\sim 10^{12}-10^{13}$ Gauss). These are relatively young neutron stars.
This is because, as a pulsar spins down (due to magnetic dipole radiation) and its 
magnetic field decays (in timescales of several million years), it eventually
ceases to be a pulsar \citep{BhattacharyaHeuvel1991}.
However, a small fraction of the radio pulsar population
is rapidly spinning (period $< 10$ ms), old (age $\sim$ billions of years),
and having weak surface magnetic fields ($10^8-10^9$ Gauss). Since many of
these pulsars are in binary systems, it has been believed for long time
that these neutron stars were spun up by the accretion-induced angular
momentum transfer in LMXB systems \citep{BhattacharyaHeuvel1991}.
This model would be supported by the observation of millisecond-period
X-ray pulsations from neutron star LMXBs, as such pulsations would be a missing link
between the millisecond radio pulsars and LMXBs. But years of 
searching did not yield the detection of any such coherent millisecond-period X-ray
pulsations. Finally in 1998, the first millisecond X-ray pulsations
(with the frequency $\approx 401$ Hz) were
discovered from the transient LMXB SAX J1808.4-3658 with the {\it RXTE}
satellite \citep{WijnandsvanderKlis1998}. As X-ray pulsations 
(like the radio pulsations) measure the neutron star spin frequency,
this showed that the neutron star in this source spins with $\approx 401$ Hz, and
hence, indeed the neutron stars in the LMXBs can be spun up. 
Several such sources have been discovered 
after SAX J1808.4-3658 (see, for example, Wijnands 2006 \citep{Wijnands2006}; 
Watts et al. 2009 \citep{Wattsetal2009}.
Fig. 10 shows the power spectrum of one such LMXB. The narrow peak 
at the frequency of 442.36 Hz shows the pulsations. Note that the computation 
of a power spectrum is a convenient way to detect timing features. Such
a spectrum (power vs. frequency) is a Fourier transform of the 
intensity profile (photon counts vs. time). For example, if $x_j
(j=0,...,N-1)$ is the number of photons detected in bin $j$, then
the discrete Fourier transform $p_k (k= -N/2,...,N/2-1)$ decomposes the
intensity into $N$ sine waves. The direct and the inverse Fourier transforms
are then given by:
\begin{eqnarray}
p_k = \sum_{j=0}^{N-1} x_j{\rm e}^{2\pi {\rm i}kj/N};
x_j = {1\over{N}}\sum_{k=-N/2}^{N/2-1} p_k{\rm e}^{-2\pi {\rm i}kj/N}.
\end{eqnarray}

%Figure got from Diego.
\begin{figure*}[h]
\includegraphics[width=7.00in]{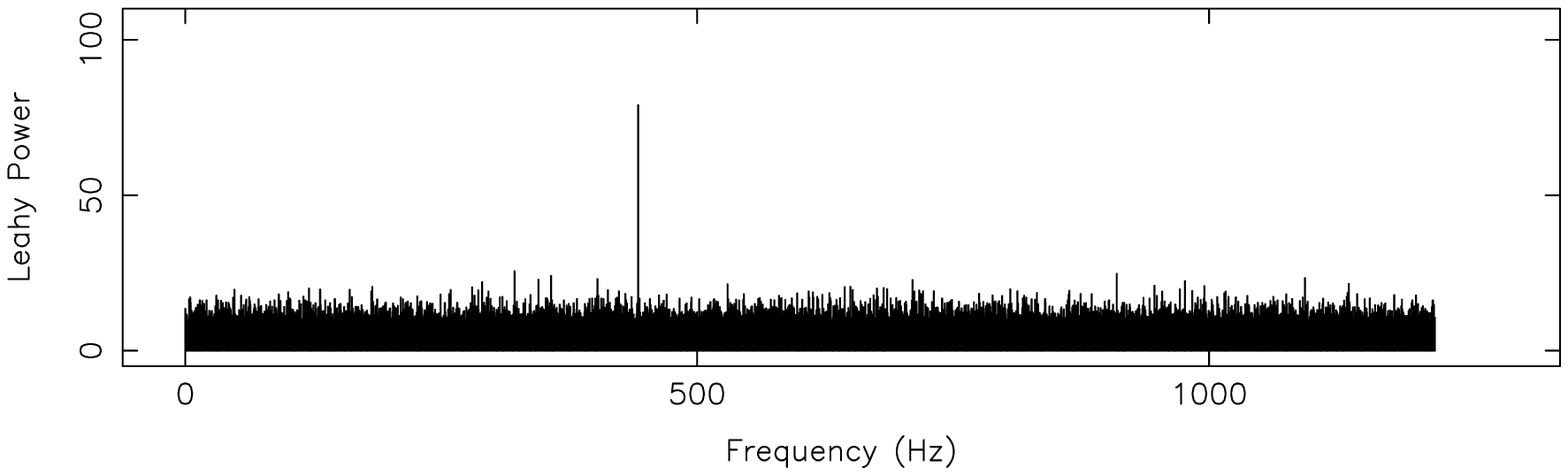}
\caption{Power spectrum (i.e., Fourier transform of the observed X-ray intensity
profile) of a neutron star LMXB. The peak at 442.36 Hz implies that (1) there
is a periodic intensity variation with that frequency; (2) this source is
an accretion-powered millisecond pulsar (see text); and (3) the spin frequency
of the neutron star is 442.36 Hz (figure courtesy: Diego Altamirano; 
Altamirano et al. \citep{Altamiranoetal2008a}).
}
\end{figure*}

In the LMXBs containing these pulsars, the accretion flow is channeled
by the neutron star magnetic field to the magnetic polar cap. The X-ray hot spot
on the neutron star created by this channeled flow rotates around the stellar
spin axis (if the magnetic axis and the spin axis are misaligned) and 
gives rise to the observed X-ray pulses.  
Since the gravitational energy release from the accreted matter powers 
the X-ray pulsations, these sources are called ``accretion-powered 
millisecond pulsars (AMPs)". Since the first discovery of these pulsars,
it intrigues the astrophysicists how these sources are different from 
other LMXBs, and why all the LMXBs do not exhibit such regular pulsations.
The recent discovery of several intermittent AMPs (that show sporadic
regular pulsations) holds the promise to resolve these problems \citep{Wattsetal2009}.
Better understanding of the AMPs will be very important to probe the
evolution of LMXBs, as well as to model the magnetically channeled
flows in the strong gravity regions near the neutron stars. Modelling
of this flow, its energy spectrum and the timing properties will be
useful to constrain the neutron star parameters, and hence the nature 
of the stellar core matter \citep{PoutanenGierlinski2003}.

\subsection{Burst Oscillations}

High frequency timing features are observed from many (so far about 20) neutron star
LMXBs during thermonuclear X-ray bursts \citep{StrohmayerBildsten2006}. 
They are called ``burst oscillations", and were first discovered with the 
{\it RXTE} PCA instrument in bursts from the LMXB 
4U 1728-34 \citep{Strohmayeretal1996}. It was soon realized that this timing feature might be 
originated from an aymmetric brightness pattern on the surface of a 
spinning neutron star. This was confirmed when burst oscillations were 
detected from the AMP SAX J1808.4-3658 close to the known neutron star
spin frequency \citep{Chakrabartyetal2003}. Later, these oscillations were
observed from other AMPs near the known stellar spin rates \citep{Strohmayeretal2003, 
Wattsetal2009}. These established a new technique to 
measure the neutron star spin frequency. In fact, regular pulsations and
burst oscillations are the only methods to measure the spin rates  
of neutron stars in LMXBs, and together they have provided spin frequencies
of about 25 sources. Apart from measuring the stellar spin, the modelling
of shapes and amplitudes of the phase-folded burst oscillation intensity 
profiles can be useful to constrain the other neutron star parameters 
(including the radius-to-mass ratio; e.g., Bhattacharyya et al. \citep{Bhattacharyyaetal2005}).
Fig. 11 shows the burst oscillation power contours and the power spectrum
for a neutron star LMXB. The power contours indicate that the oscillation 
frequency changes with time. Indeed such small evolution of frequencies
are observed for many bursts, which suggests that the aymmetric brightness 
pattern slowly moves on the spinning neutron star surfaces. It is not
yet understood what causes such motion, although several models exist
in the literature \citep{StrohmayerBildsten2006}. The oscillations
during burst rise are normally believed to be caused by the expansion
of the burning region on the stellar surface immediately after the 
ignition. Therefore, the study of the evolution of the burst oscillation 
properties during burst rise can be useful to probe the thermonuclear
flame spreading. The nature of the aymmetric brightness pattern that 
causes the oscillations during burst decay is not yet well understood.
This is because the entire stellar surface is believed to be ignited
at this time. The existing models involve non-radial global oscillations
in the surface layers \citep{Heyl2004}, shear oscillations \citep{Cumming2005},
etc., each of which has shortcomings.

%Updated figure of A1744-361 given.
\begin{figure}[btp]
\includegraphics[width=3.00in]{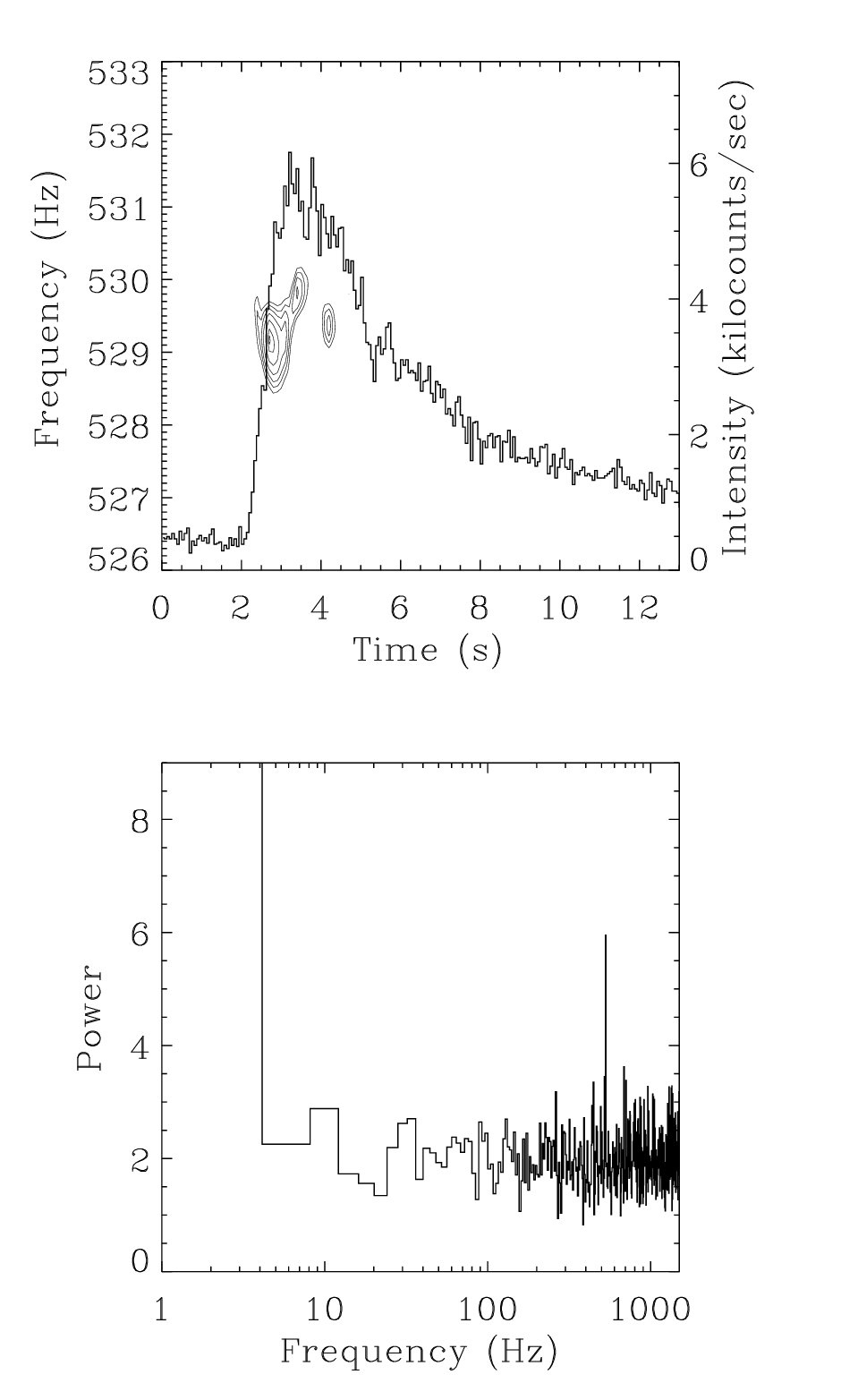}
\caption{Upper panel: X-ray intensity profile of a thermonuclear X-ray burst
from the neutron star LMXB 1A 1744-361 \citep{Bhattacharyyaetal2006b}.
Power contours around 530 Hz (calculated by Fourier transform) implies
that there is a significant intensity variation (burst oscillation) with $\approx 530$ Hz
frequency. Note that the shape of the power contours indicates a slight change
in frequency with time, which is common for burst oscillation. 
Lower panel: The corresponding power spectrum. The
peak at $\approx 530$ Hz implies the burst oscillation with that frequency.
}
\end{figure}

\subsection{Quasi-periodic Oscillations and Broadband Power Spectrum}

%Updated version of kHz QPOs from 4U 1636-536 made.
\begin{figure}
\includegraphics[width=3.00in]{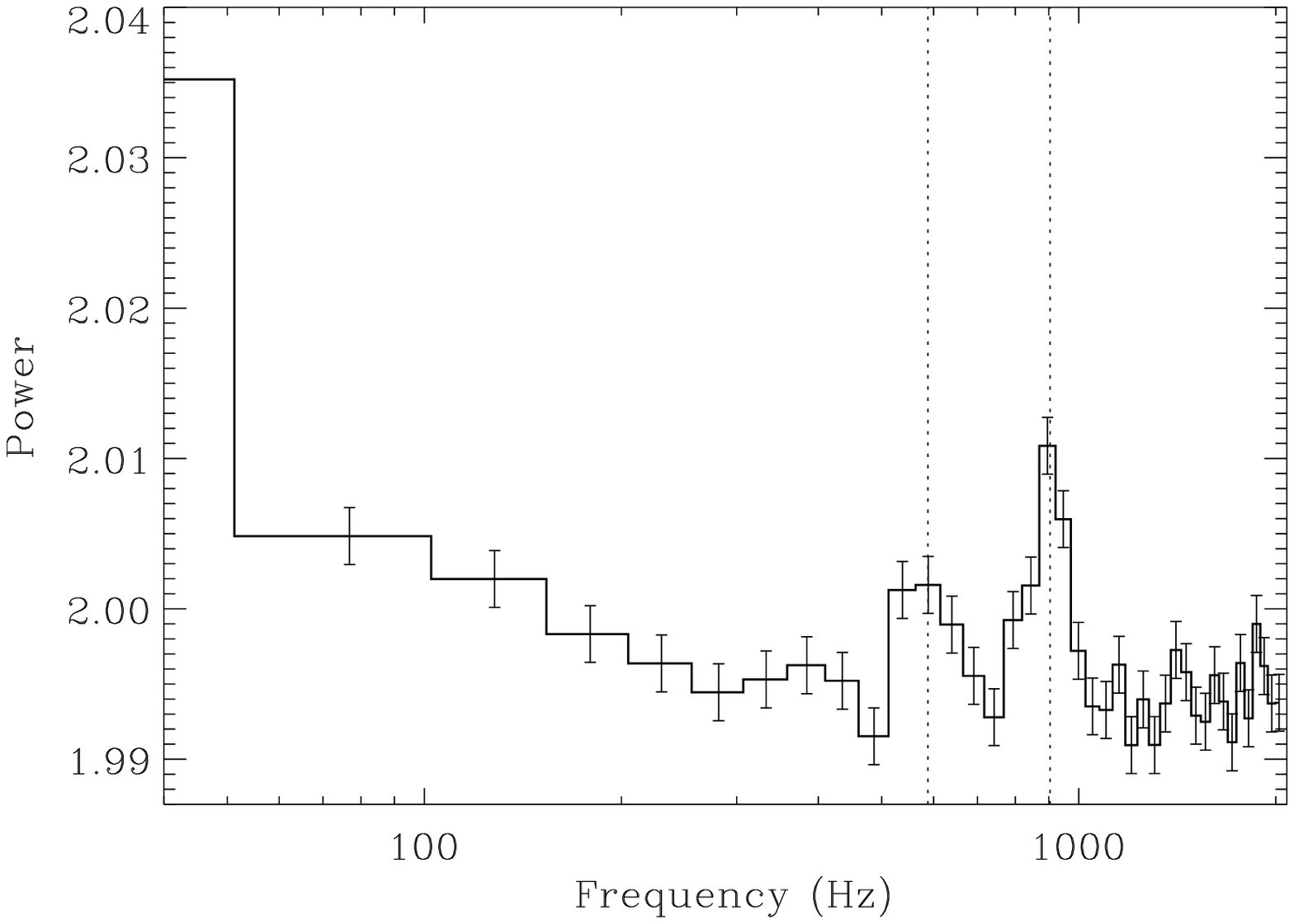}
\caption{Power spectrum (i.e., Fourier transform of the observed X-ray intensity
profile) of a neutron star LMXB showing a pair of kHz QPOs (see text).
The vertical dotted lines show the centroid frequencies of these QPOs.
}
\end{figure}

Neutron star LMXBs show a variety of observationally robust timing features, 
many of which are not well understood yet. Continuum power, in excess of the
mean random power, is often observed at frequencies less than 100 Hz. 
The nature of this power depends on the source state determined from the
position on the color-color plot. For example, a power-law-like continuum
($\propto \nu^{-\alpha}$; $\nu$ is the frequency in the power spectrum; and $\alpha$
is a constant) below 1 Hz is typically dominant for the atoll sources in
the end part of the banana state \citep{vanderKlis2006}. Apart from the broad
continua, neutron star LMXBs often exhibit peaked features with a wide range
of frequencies (millihertz to kilohertz). Since they are somewhat broad, and 
hence not quite periodic, these features are known as quasi-periodic 
oscillations (QPOs). Similar to the continuum features, QPOs are correlated
with the color-color diagram. For example, a type of $15-60$ Hz QPO appears in the 
horizontal branch of the color-color diagram of Z sources, and hence known as the
``horizontal branch oscillation (HBO)". The QPOs, when modelled correctly, 
are expected to significantly enhance our understanding of neutron star LMXBs.
The observed correlations among different types of QPOs can be helpful for their
modeling. However, in this short review we will not discuss the QPOs and
their correlations in detail, and we refer to an extensive review by 
van der Klis \citep{vanderKlis2006} that gives a large amount of information about the fast 
X-ray timing properties of neutron star LMXBs. 
The only QPOs, that we would like to briefly discuss here, are called 
kilohertz (kHz) QPOs. These high frequency QPOs are observed from both Z and atoll sources.
The frequencies of kHz QPOs vary between $\approx 200-1200$ Hz, and often they appear
in a pair (see Fig. 12). For a given source, the lower and upper kHz QPOs are seen to move up and
down in frequency together without changing the frequency separation much.
This frequency separation is normally found to be within 20\% of the neutron star
spin frequency, or half of that, depending on source. These kHz QPOs
are believed to be originated from a region close to the neutron star, and
hence, when their origin is fully understood, will be very useful to measure
the neutron star parameters, and to test the theory of general relativity
(e.g., Miller et al. 1998 \citep{Milleretal1998}; Stella \& Vietri 
1998 \citep{StellaVietri1998}; Barret et al. 2005 \citep{Barretetal2005}).

\section{Future Prospects}

%Figure given by Didier.
\begin{figure}
\includegraphics[width=3.00in]{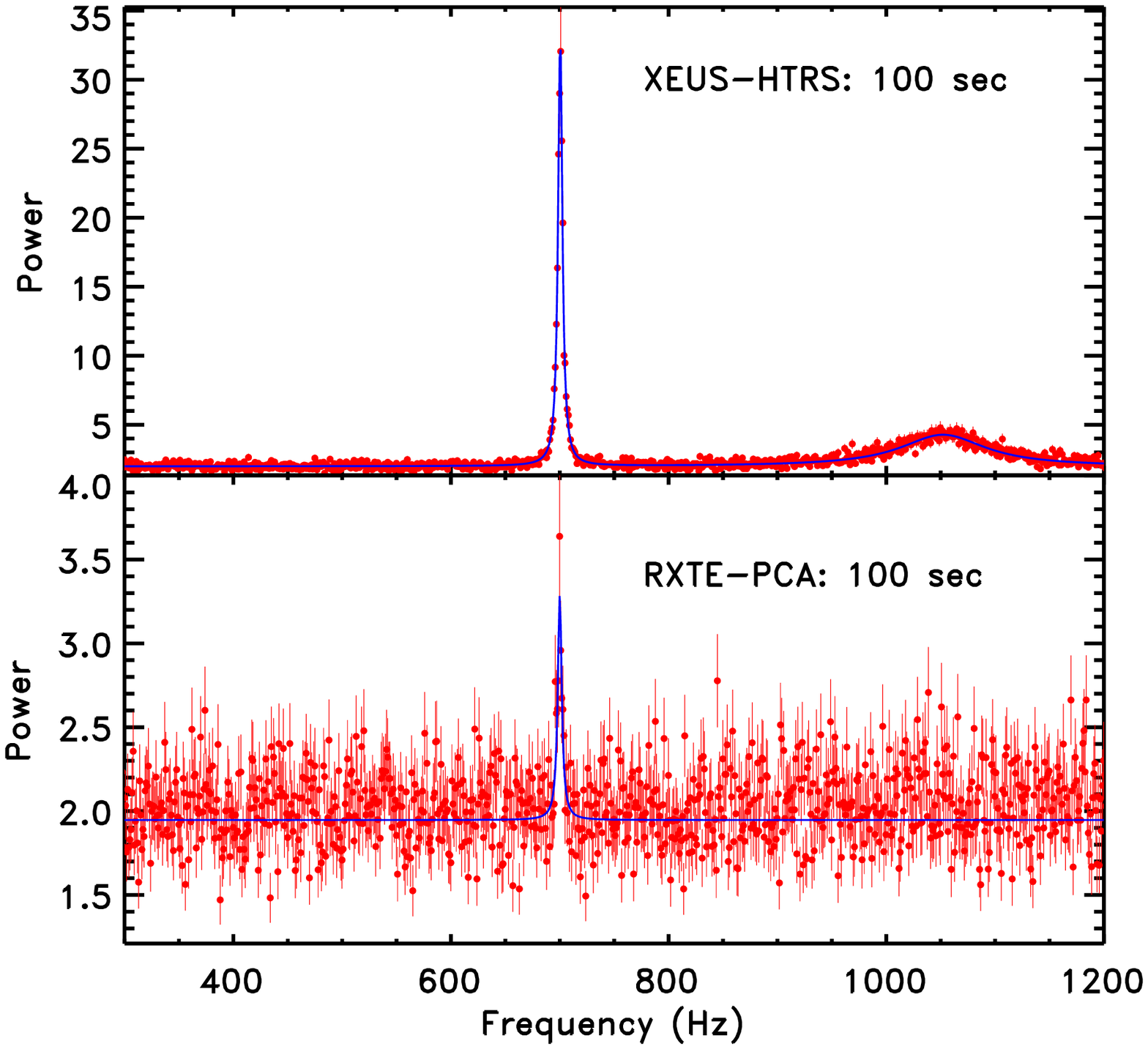}
\caption{A comparison of a power spectrum observed with two (one current and one future)
instruments. The lower panel shows that the lower kHz QPO is barely detected and
the upper kHz QPO is not detected from the power spectrum created from an intensity
profile of 100 seconds duration observed with the PCA instrument of the {\it RXTE}
satellite. The upper panel shows that, if the same observation were made with the
High Time Resolution Spectrometer (previously proposed for the {\it X-Ray Evolving 
Universe Spectrometer} ({\it XEUS}), but now proposed for the {\it International 
X-ray Observatory} ({\it IXO})), both the kHz QPOs would be very significantly 
detected (figure courtesy: Didier Barret; Barret et al. \citep{Barretetal2008}).
}
\end{figure}

Neutron star LMXBs provide a unique opportunity to study several aspects
of extreme physics that we have briefly discussed in this short review.
These include (1) constraining the equation of state models of the
supranuclear density matter of neutron star cores from the measurements
of the stellar parameters;
(2) testing the theory of general relativity; (3) understanding the
physics of neutron star atmospheres; and (4) probing the accretion flow
in the strong gravity region near the neutron stars. In addition, the X-ray
study of neutron star LMXBs allows us (1) to understand the evolution
of these binary systems; (2) to probe the long term accretion process and
the structures of the accreted matter; and (3) to find out how and why 
these systems are different from black hole LMXBs.
Since the observed X-ray properties mentioned in this review can be
used as tools to achieve these scientific goals, we need to understand
these properties better, and explore (1) the correlations among them, and (2)
their correlations with the observations in other wavelengths.
These can be achieved with more sensitive X-ray instruments (typically
with larger photon collecting areas), and simultaneous observations
in multiple wavelengths. It will also be required to monitor the entire
sky frequently with high sensitivity in order to catch the transient 
neutron star LMXBs during their outbursts. Here we give a short
description of some of the proposed future space missions with these
capabilities.

The proposed Indian multiwavelength astronomy space mission {\it Astrosat}
will observe simultaneously in a wide energy range (optical to hard X-ray of
100 keV). Apart from this unprecedented capability, its LAXPC instrument
should be able to detect and measure the high frequency timing features,
such as kHz QPOs and regular pulsations, for the first time in hard X-rays
(say, up to $\approx 50$ keV). Note that in its own time, LAXPC will
probably be the only instrument capable of detecting the high frequency
features in the X-ray band. 

The proposed next generation joint American, European and Japanese satellite 
{\it International X-ray Observatory} ({\it IXO}) will have several 
highly sensitive instruments that will take the X-ray study of neutron star LMXBs
to a higher level. In Fig. 13, we demonstrate how significant an improvement
the high timing resolution spectrometer (HTRS) instrument of {\it IXO} 
will make over the {\it RXTE} PCA for the high frequency timing observations.
Moreover, Ray et al. \citep{Rayetal2009} mention that the proposed {\it Advanced X-ray 
Timing Array} ({\it AXTAR}) satellite will have better capabilities than
{\it IXO} for these timing observations. Finally, the Japanese mission
{\it Monitor of All-sky X-ray Image} ({\it MAXI}) will monitor the
X-ray sky in the $0.5-30$ keV range with high sensitivity. Therefore,
the study of the neutron star LMXBs is likely to remain an active and
promising field in order to explore some aspects of extreme physics.

{}

\end{document}